\documentclass[aps,prl,twocolumn,superscriptaddress]{revtex4-1}

\usepackage{graphicx}
\usepackage{latexsym}
\usepackage{amsmath}
\usepackage{amssymb}
\usepackage{amsfonts}
\usepackage{color}

\begin{document}

%\preprint{}

\title{Vortex core deformation and stepper motor behavior in a superconducting ratchet}

\author{J. Van de Vondel}
\affiliation{INPAC, Katholieke Universiteit Leuven, Celestijnenlaan
200D, B--3001 Leuven, Belgium}
\author{V. N. Gladilin}
\affiliation{INPAC, Katholieke Universiteit Leuven, Celestijnenlaan
200D, B--3001 Leuven, Belgium}
\affiliation{TQC, Universiteit
Antwerpen, Universiteitsplein 1, B-2610 Antwerpen, Belgium}
\author{A.V. Silhanek}
\affiliation{INPAC, Katholieke Universiteit Leuven, Celestijnenlaan
200D, B--3001 Leuven, Belgium}
\author{W. Gillijns}
\affiliation{INPAC, Katholieke Universiteit Leuven, Celestijnenlaan
200D, B--3001 Leuven, Belgium}
\author{J. Tempere}
\affiliation{TQC, Universiteit Antwerpen, Universiteitsplein 1,
B-2610 Antwerpen, Belgium}
\author{J. T. Devreese}
\affiliation{TQC, Universiteit Antwerpen, Universiteitsplein 1,
B-2610 Antwerpen, Belgium}
\author{V. V. Moshchalkov}
\affiliation{INPAC, Katholieke Universiteit Leuven, Celestijnenlaan
200D, B--3001 Leuven, Belgium}

\date{\today}

\begin{abstract}
We investigated experimentally the frequency dependence of a
superconducting vortex ratchet effect by means of electrical
transport measurements and modeled it theoretically using the time
dependent Ginzburg-Landau formalism. We demonstrate that the high
frequency vortex behavior can be described as a discrete motion of a
particle in a periodic potential, i.e. the so called stepper motor
behavior. Strikingly, in the more conventional low frequency
response a transition takes place from an Abrikosov vortex rectifier
to a phase slip line rectifier. This transition is characterized by
a strong increase in the rectified voltage and the appearance of a
pronounced hysteretic behavior.

%
%The obtained results revealed a low frequency rectification ruled
%by kinematic vortices or phase slip lines which evolves to a
%particle rectifier in the high frequency range. The simulations
%indicate that the presence of these kinematic vortices is
%responsible for an apparent mass ruling their dynamics. Moreover,
%at high frequency drives, we observed the stepper motor behavior,
%which is a discrete motion of the vortex lattice in the periodic
%pinning potential.
\end{abstract}

\pacs{}
%\keywords{}

\maketitle


%\section{Introduction}

Since the early 90's, inspired by the pioneering work of
Magnasco~\cite{Magnasco93}, a strong theoretical effort has been
devoted to modeling complex biological ratchet systems using the
one-dimensional overdamped Langevin equation of motion for a single
particle~\cite{Ajdari94, marchesoni}. In this approximation, the
possibility of a directional motion was demonstrated for a Brownian
particle, affected by an external time-correlated driving force and
a spatially periodic potential, which lacks reflection symmetry. A
common prediction of these models is discretization of the particle
dynamics under a periodic driving force: the average velocity of the
particles can only acquire discrete values corresponding to an
integer number of unit cells traveled by the particles during each
cycle of the drive. Soon after these theoretical predictions were
published, the main concepts have been successfully applied in
experimental research both at microscopic and macroscopic
scales~\cite{marchesoni,Linke06}. A particularly ideal playground
with a high level of flexibility allowing to explore the influence
of particle density, particle size, and to cover the whole space of
parameters of the ratchet potentials is provided by vortices in a
type-II superconducting film~\cite{Lee99}. Here, the particle is a
quantized flux bundle (vortex), which is induced by an external
magnetic field and which has a temperature-dependent size. These
vortices can be driven by an applied electrical current flowing
through the superconductor whereas the periodic asymmetric potential
to trap the vortices can be generated by using modern lithographic
techniques~\cite{Villegas03,VandeVondel05}.

However, two important features inherent to vortex ratchets have not
been properly taken into account by the current theoretical models.
Firstly, the repulsive interaction between vortices makes their
dynamics more complex. Indeed, the competition between the
asymmetric pinning and the vortex-vortex repulsion leads to
reversals of the preferential flow direction of vortices as their
density increases~\cite{Derenyi95, deSouzaSilva06}. Secondly, the
one-to-one mapping of a vortex to a particle breaks down at high
vortex velocities where vortex displacements take place at shorter
time scales than the healing time to recover the superconducting
order parameter along the vortex trajectory. As the vortex velocity
is progressively increased, this latter effect leads first to a
dynamic reconfiguration of the vortex lattice, followed by the
elongated-vortex motion which is much faster than that of the
Abrikosov vortices. The driven elongated vortices are eventually
transformed into phase slip lines~\cite{Larkin75, Vodolazov07}. This
bring us to the formulation of the main unsolved problems that will
be addressed in the present work, (i) how do the internal degrees of
freedom of a vortex influence the rectification properties of vortex
ratchets? (ii) is it possible to realize the theoretically predicted
but so far experimentally not observed vortex stepper motor
behavior?

%
%

%\section{Experimental details and the model}

\begin{figure}[h]
\centering
\includegraphics*[angle=0, width=1\linewidth]{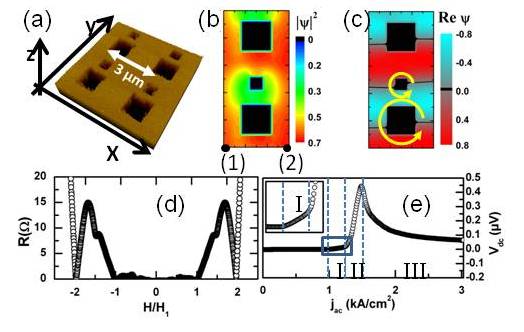}
\caption{ (a) Atomic Force Microscopy image of a 5 $\times$ 5 $\mu$m
area of the sample. (b) $|\psi|^{2}$ and (c) Re $\psi$ calculated at
$H = 2H_1$ using the TDGL model, with $H_1$ the magnetic field
generating exactly one single-quantum vortex in each unit cell.
Experimentally obtained: (d) resistance versus field measured at 1.2
K, and with dc current of 100 $\mu$A. (e) The dc voltage ($V_{dc}$),
across a single row of antidots, as a function of the driving
amplitude with a frequency of 100 kHz. As explained in the text,
three different dynamical regimes can be identified (labeled I, II
and III) as a function of the ac amplitude. The zoom in panel (e)
gives the voltage dependence in region I.\label{Figure1}}
\end{figure}

\emph{Experimental details and theoretical model.-}In order to
address these important issues we have investigated a
superconducting vortex rectifier consisting of a 33 nm thick Al
bridge patterned with a periodic array of asymmetric pinning sites
with a unit cell consisting of two square antidots of different
sizes (1200 nm and 600 nm), separated by 400 nm distance [See Fig.
\ref{Figure1}(a)]. The superconducting critical temperature is
$T_{c}$=1.445 K. The coherence length and the penetration depth are
$\xi(0) \sim 73$~nm and $\lambda(0) \sim 261$~nm, respectively. When
applying an ac current along the bridge ($x$-axis), the vortices,
induced by a magnetic field perpendicular to the plane of the bridge
($z$-axis), are subjected to a driving force in the $y$-direction.
The resulting motion of vortices leads to dissipation, which is
probed by measuring the voltage drop along the bridge. The
superconducting vortex rectifier has been investigated both
experimentally in a $^{4}$He cryostat and theoretically using
time-dependent Ginzburg-Landau (TDGL) simulations. The previously
used theoretical approaches (Langevin dynamics, Fokker-Planck
equations, molecular dynamics), which are not specific for
superconducting systems, deal with motion of {\it rigid entities}
embedded in a medium, while vortices in a superconductor correspond
to ``soft'' superconducting current loops which can be deformed by
applying exteranl drive suchas the Lorentz force. Unlike these
approaches, the TDGL formalism~\cite{gorkov,Kramer78} treats the
dynamics of the condensate as a whole, in terms of the
time-dependent complex order parameter $\psi({\bf r}, t)$. This
treatment contains all necessary ingredients, including vortex
deformability, to reproduce and explain the experimental data.

To simplify the presentation, we concentrate on the experimental
data obtained at the temperature $T=$1.430 K and in a particular
magnetic field $H$, which generates one flux quantum per each
antidot. The equilibrium vortex configurations are visualized by
plotting the square modulus of the order parameter, $|\psi|^{2}$
[Fig. \ref{Figure1}(b)], and its real part, \mbox{Re $\psi$} [Fig.
\ref{Figure1}(c)]. Since \mbox{Re $\psi$} should change sign two
times along a closed path encircling a single flux quantum,
figure~\ref{Figure1}(c) indicates the presence of one
$\Phi_0$-vortex trapped by each antidot. The stability of this
vortex configuration leads to a minimum in dissipation (i.e.
resistance) at this field~\cite{baert95}, as shown in Fig.
\ref{Figure1}(d).

\emph{Influence of the internal degrees of freedom .-} Figure
\ref{Figure1}(e) shows the measured rectified (dc) voltage $V_{dc}$
(proportional to the average vortex velocity) as a function of the
amplitude of a sinusoidal excitation at a low frequency $f=100$~kHz.
The positive dc voltage sign corresponds to particles moving
preferentially in the $-y$ direction. This motion, opposite to the
preferential direction at $H=H_1$, is caused by the competition
between the vortex-vortex repulsion and the asymmetric pinning
landscape~\cite{deSouzaSilva06}. A closer look at the amplitude
dependence of $V_{dc}$ reveals three distinct regions of
rectification. At low ac amplitudes $j_{ac}$, region I, a linear
increase of $V_{dc}$ is present (a close-up in the inset makes
region I more visible). With increasing ac amplitude this develops
into a much steeper, non-linear increase in region II. The maximum
rectification accounts for the onset of region III, where a sharp
decrease of $V_{dc}(j_{ac})$ takes place. A single-particle model,
as used in Ref.~\cite{Magnasco93,Ajdari94}, cannot describe the
change in the rectifying properties between region I and II. As we
demonstrate below, the sudden increase in the measured dc voltage
results from a transformation of the moving vortices and their
environment.

\begin{figure}[h]
\centering
\includegraphics*[angle=0, width=1\linewidth]{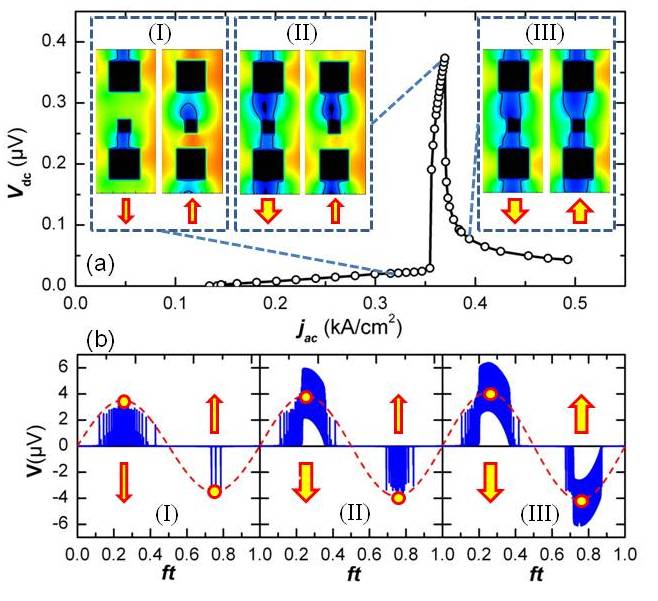}
\caption{\label{Figure2} (a) Calculated dc voltage, V$_{dc}$, as a
function of the amplitude of the ac current density for $\gamma =
10$ and $f = 400$~kHz. (b) Voltage drop between points (1) and (2)
shown in Fig. 1(b), as a function of time for different ac currents:
(I) $j_{ac}=0.33$~kA/cm$^2$, (II) $j_{ac}=0.37$~kA/cm$^2$, (III)
$j_{ac}=0.39$~kA/cm$^2$. The dashed curve in panel (b) shows the
dynamics of the applied current density. For the same ac currents,
snapshots of $|\psi|^{2}$ are shown in the insets of (a), labeled
correspondingly (I),(II),(III). The snapshots were taken during the
ac cycle at maximum applied positive and negative currents. The thin
and thick arrows indicate the direction of motion for Abrikosov and
kinematic vortices, respectively.}
\end{figure}

In order to elucidate the importance of the internal degrees of
freedom of the vortices and the superconducting condensate for the
dc voltage read-out, we performed simulations using the generalized
TDGL equations~\cite{Kramer78}. An important model parameter,
influencing the internal structure of a moving vortex, is
$\gamma=8k_BT_c\tau_{in}\sqrt{u(1-T/T_c)}/(\pi\hbar)$, with
$\tau_{in}$, the inelastic collision time for electron-phonon
scattering, and $u\approx 5.79$. At $\gamma \gg 1$ the healing time
of the condensate is $\tau\sim \gamma |\psi| \tau_{GL}$, where
$\tau_{GL}$ is the Ginzburg-Landau relaxation
time~\cite{Vodolazov07}. In our calculations we have checked the
behavior of the system for a wide range of $\gamma$ values. All
other model parameters coincide with the experimental values.

The dependence of $V_{dc}$ on the driving amplitude $j_{ac}$ is
shown in Fig. \ref{Figure2}(a) for $f= 400$~kHz and $\gamma = 10$.
The TDGL simulations reproduce the qualitative features of the
experimental data [Fig. \ref{Figure1}(e)], showing the same three
distinct regions of rectification. Between 0.14 kA/cm$^{2}$ and 0.37
kA/cm$^{2}$ an increase of the dc voltage is observed with a
discontinuity at 0.36 kA/cm$^{2}$. At this point the dependence of
the dc voltage on $j_{ac}$ changes from an approximately linear
increase (region I, 0.14 kA/cm$^{2}$ $< j_{ac}<$ 0.36 kA/cm$^{2}$)
to a steeper non-linear increase (region II, 0.36 kA/cm$^{2}$ $<
j_{ac} <$ 0.37 kA/cm$^{2}$).

In order to reveal the origin of these features in $V_{dc}(j_{ac})$
we investigated the time evolution of the voltage difference across
a single row of antidots [between points (1) and (2) in Fig.
\ref{Figure1}(b)]. At 0.33 kA/cm$^{2}$ (region I) a clearly
asymmetric motion of individual vortices is observed. A multitude of
peaks in the voltage [Fig. \ref{Figure2}(b)I] during one ac cycle
indicates the passage of a series of vortices. A snapshot of
$|\psi|^{2}$ [Fig. \ref{Figure2}(a), Inset I] at maximum applied
positive and negative currents indicates the presence of a vortex
with a localized normal core at these low current drives. Since the
dynamics of the condensate is limited by a specific time scale
$\tau$, a drastic change occurs if the condensate does not have
enough time to fully recover before the passage of a new vortex
\cite{Vodolazov07}. Under these circumstances the vortex will be
deformed at higher velocities and a phase slip line (PSL) or
kinematic vortex appears. A snapshot of $|\psi|^{2}$ in region II at
$j_{ac}=0.37$~kA/cm$^{2}$ [Fig. \ref{Figure2}(a), Inset II] indeed
shows the depletion of the order parameter along the direction of
the vortex motion. In region II the broken symmetry of the vortex
motion results in the generation of such a PSL for positive applied
currents only. The transformation between the motion of vortices
with a localized core to a PSL exactly coincides with the boundary
between the two different regimes of rectification. The motion of
kinematic vortices is much faster and results in a highly increased
voltage during the positive half-period of the ac drive [Fig.
\ref{Figure2}(b)II] and, as such, it explains the drastic increase
of $V_{dc}$  in region II. Above 0.37 kA/cm$^{2}$ (region III) a PSL
is created during both positive and negative half-periods of the
applied current [panel (b)III and Inset III in Fig.~\ref{Figure2}],
thus giving rise to a strong reduction of the average voltage signal
at higher ac drives. It is important to note that in the PSL-like
regime the vortex core elongates and it is impossible to see a
vortex as a single entity. In this regime, the representation of
this system by a point particle model in a periodic potential breaks
down. Instead, a proper treatment of the vortex within the TDGL
formalism is needed to explain the observed ratchet effects.

\begin{figure}[h!]
\centering
\includegraphics*[angle=0, width=1\linewidth]{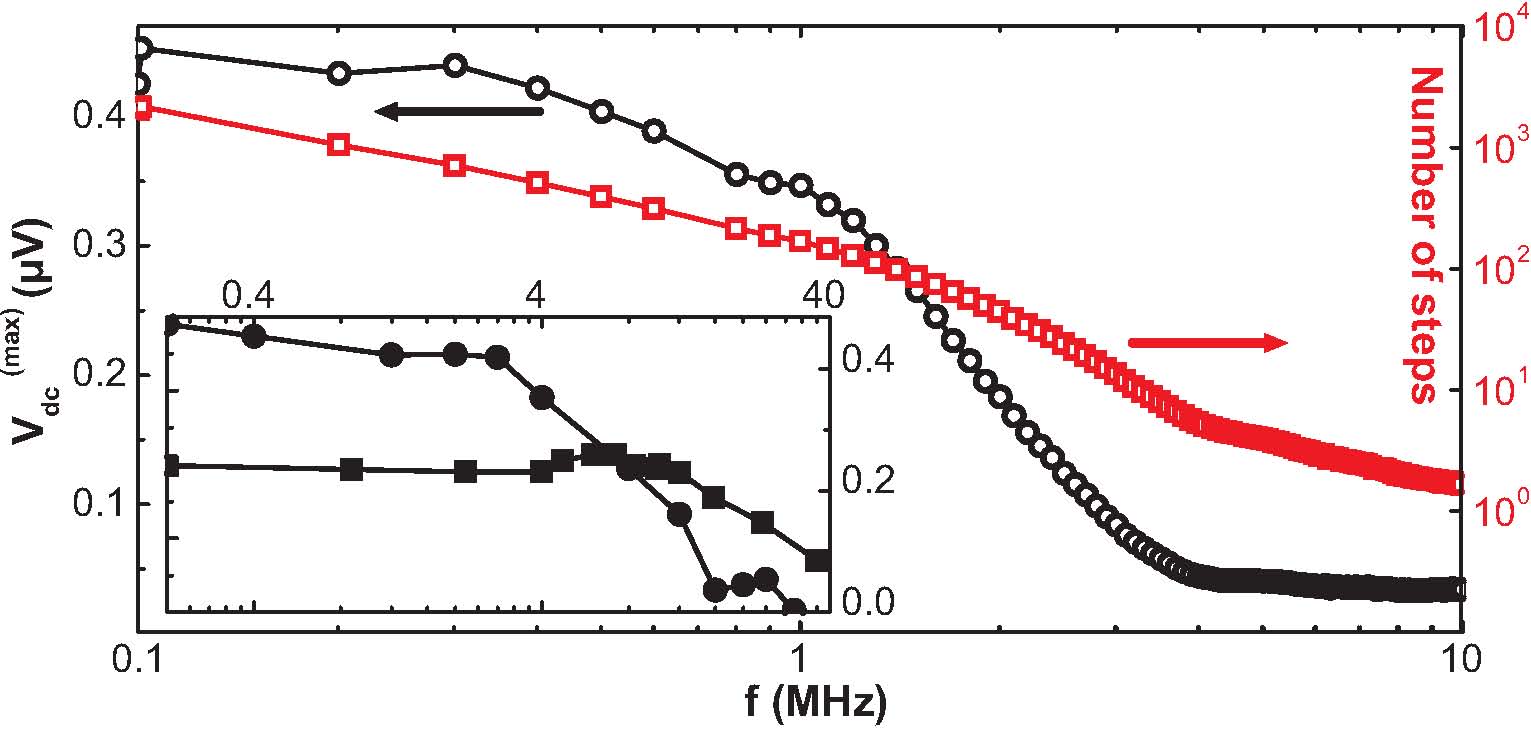}
\caption{Maximum rectification (open circles) and the number of
steps (open squares) as a function of the frequency $f$ at $T =
1.430$~K and $H = 2H_{1}$. Inset: Maximum rectification,
$V_{dc}^{\rm (max)}$ ($\mu$V), calculated within the TDGL model for
$\gamma = 5$ (full squares) and 10 (full circles), as a function of
$f$ (MHz). \label{Figure3}}
\end{figure}

\emph{Frequency dependence.-} Since the transition from Abrikosov
vortices to PSL is related to $\tau$, a change in the rectification
properties is expected if the period of the driving force becomes
comparable to this characteristic time scale. Unambiguous evidence
of this change comes from Fig.~\ref{Figure3} where we plotted the
maximum rectified voltage, $V_{dc}^{\rm (max)}$, as a function of
the applied ac frequency (open circles). Above 300 kHz a strong
decrease of rectification is observed until 3Mhz, beyond which only
a small decay of the rectification signal is found.

As discussed in the previous section, at low frequency drives, the
maximum rectification is determined by the unidirectional creation
of a PSL. The onset of PSL corresponds to the regime in which the
condensate is incapable of full recovery in between two consecutive
vortex passages. However, if the period of the driving force becomes
comparable to the lifetime of a PSL, the PSL regime persists for
both current directions. In the experiment this occurs above 3 Mhz.
As a result, the PSL in this regime does not significantly
contribute to the rectification signal (region II disappears) and a
strong reduction of the maximum rectification is expected. For those
frequencies, the maximum of V$_{dc}$ is achieved just before the
onset of PSL, when rectification is dominated by the ``conventional
vortex hopping''. A qualitatively similar frequency dependence is
obtained in TDGL simulations [Inset of Fig.~\ref{Figure3}],
indicating that the model captures the main physical mechanism. A
comparison between the different $\gamma$ values [$\gamma=10$ (full
circles) and 5 (full squares)] shows a shift of the overall behavior
towards lower frequencies at higher $\gamma$ values. This effect is
explained by the healing time of the condensate $\tau$ being
proportional to $\gamma$, so that the condensate acts slower for
larger $\gamma$ values. In line with this, the relatively low
frequency scale in the experiment can be attributed to the fact that
the $\gamma$ value experimentally obtained for Al~\cite{Lawrence78}
is $\sim 500$.

\emph{Discrete rectification.-} We concluded above that at
sufficiently high frequencies the rectification is governed by the
directional motion of Abrikosov vortices. In this regime a shorter
timescale appears, associated with the traveling time of a vortex
from cell to cell. The synchronization between an external
excitation and the internal natural frequency should lead to a
discrete hopping of vortices known as a vortex stepper motor. The
average voltage generated by a single moving vortex row is given by
$V_{dc} = n_{step}\Phi_0 f$ with $\Phi_0=2.07 \times 10^{-15}$
T\,m$^2$ the flux quantum, $f$ the frequency in Hz, $n_{step}$ the
number of steps taken by the vortex row. Knowing $f$, we can
calculate the number of the unit cells the vortex is moving during a
single ac cycle (as shown in Fig.~\ref{Figure3}, red circles). At
low driving frequencies, this number is large ($n_{step}> 100$). As
a result, any discreteness of the vortex motion becomes difficult to
resolve. However, above 3 MHz the number of steps taken per cycle is
less than 10 and discrete behavior of the vortex lattice should
appear.

\begin{figure}[h!]
\centering
\includegraphics*[angle=0, width=0.85\linewidth]{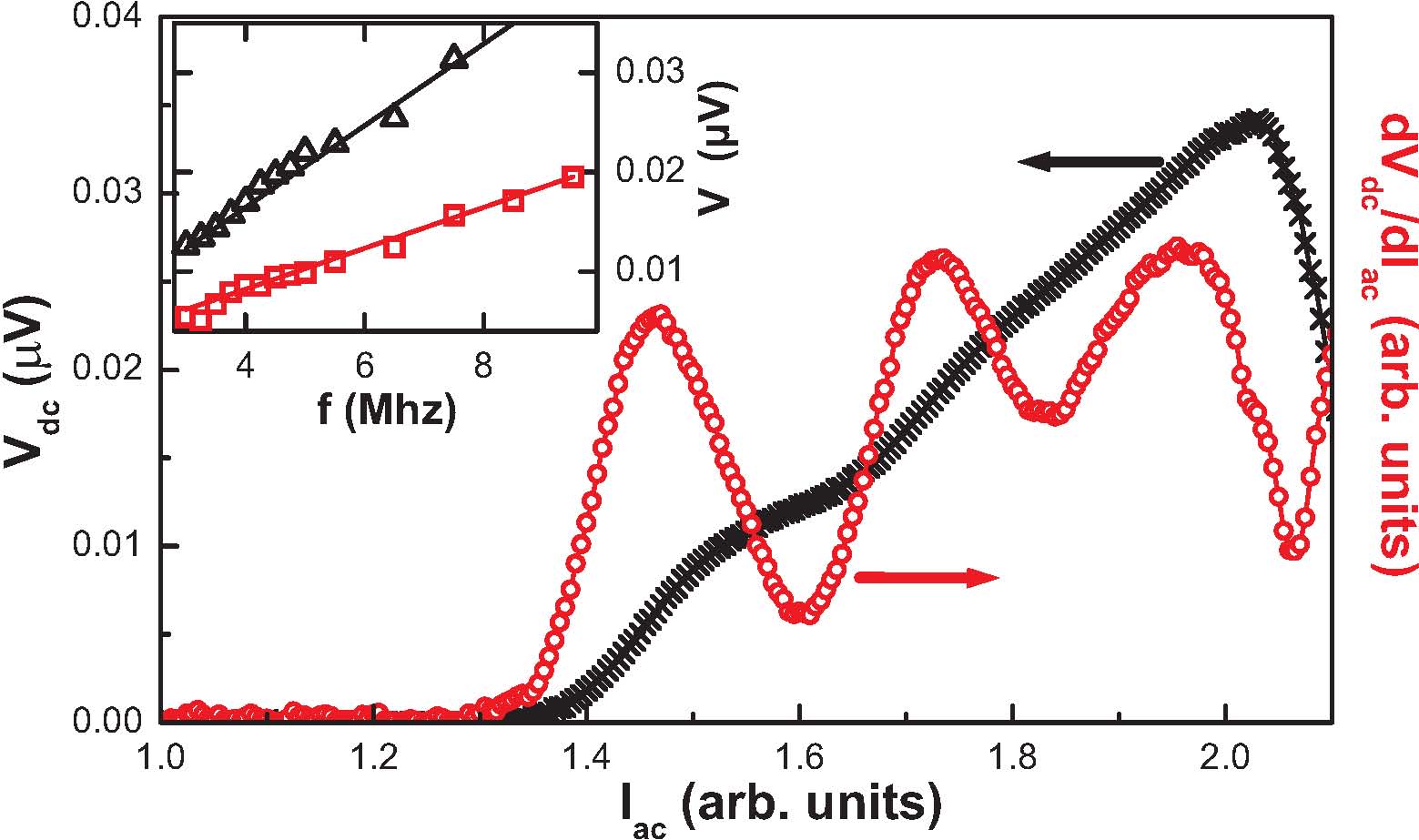}
\caption{$V_{dc}$ (crosses) and  $\frac{dV_{dc}}{dI_{ac}}$ (circles)
as a function of $I_{ac}$ with $f$ = 6.5 MHz. Inset: Height of the
first (squares) and second (triangles) voltage step as a function of
$f$. The full lines indicate the expected values $V_{dc} =
n_{step}\Phi_0 f$. \label{Figure4}}
\end{figure}

Figure \ref{Figure4} shows a representative $V_{dc}(I_{ac}$) curve
(crosses) in the high frequency regime ($f$ = 6.5 MHz). Here a
smoothed staircase structure is observed. The fact that the discrete
hopping of vortices does not manifest itself as very sharp voltage
steps, can be explained by the large number of antidot rows (333)
which contribute to the measured signal. This could lead to
statistical spreading. A common practice to unveil these blurred
steps is to measure the derivative $\frac{dV_{dc}}{dI_{ac}}$, as
shown in Fig.~\ref{Figure4} (circles). The minima in
$\frac{dV_{dc}}{dI_{ac}}$ are used to determine the locations of the
steps on the current axis and the height of the steps is defined as
the voltage output exactly at this location. In the inset of
Fig.~\ref{Figure4} we plotted the height of first (squares) and
second (triangles) voltage step together with the theoretical
prediction  $V_{dc} = n_{step}\Phi_0 f$ (red and black lines). The
nice overlap between experiment and theory demonstrates that we
indeed achieved a coherent stepper motor behavior of the whole
Abrikosov lattice in the periodic pinning landscape.

To conclude, the main points of this letter are (i) the realization
of the vortex stepper motor, particularly evident at high frequency
excitations, and (ii) the presence of a driving force window where
the nature of the moving particle changes from an Abrikosov vortex
to a phase slip line, when switching the direction of the driving
force. The former effect is very sensitive to the particular value
of the magnetic field, whereas the latter is a more robust effect.
The formation of phase slip lines gives rise to the experimentally
observed irreversibilities which have been previously explained by
attributing an artificial mass to non-deformable
vortices~\cite{VandeVondel05}. We believe that our results have
relevance for other ratchet systems of deformable objects or
media~\cite{Linke06}.

This work was supported by Methusalem funding by the Flemish
government, the Flemish Science Foundation (FWO-Vl), in particular
FWO projects G.0356.05, G.0115.06, and G.0370.09N, the Belgian
Science Policy, and the ESF NES network. A.V.S., J.V.d.V and W.G.
acknowledge support from FWO-Vl.


\end{document}